# For optical flickering in symbiotic star MWC 560

**Dragomir Marchev[1], Kiril Stoyanov[2], Vladislav Marchev[1,2], Radoslav Zamanov[2], Borislav Borisov[1], Doroteq Vasileva[1], Teodora Atanasova[1], Nataliya Pavlova[3]**

[1] *University of Shumen, Faculty of Natural Sciences, Department of Physics and Astronomy,*
*115 Universitetska Str., 9700 Shumen, Bulgaria,*
*E-mail: d.marchev@shu.bg, b.borisov@shu.bg, d.vasileva@shu.bg, t.atanasova@shu.bg*

[2] *Institute of Astronomy and National Astronomical Observatory, Bulgarian Academy of Sciences,*
*72 Tsarigradsko Shose Blvd., 1784 Sofia, Bulgaria,*
*E-mail: kstoyanov@astro.bas.bg, vmarchev@astro.bas.bg, rkz@astro.bas.bg*

[3] *University of Shumen, Faculty of Mathematics and Informatics, Department of Algebra and Geometry,*
*115 Universitetska Str., 9700 Shumen, Bulgaria,* E-mail: n.pavlova@shu.bg

**Abstract:** *This study is based on observations of MWC560 during the last two observational seasons (2020/2021 and 2021/2022). Other than looking for flickering we were interested in following the variability of brightness in the same period. Looking for similarities in the spectra with other types of stars is also of great interest to us because it could help clarify the stellar configuration of such objects. Our observations during the last two observational seasons of MWC560 confirm the absence of flickering. From the similarities of the gathered spectra of XX Oph and MWC560 we assume that the components in XX Oph are a red giant and a white dwarf, which are also surrounded by a common shell.*

**Introduction**

The symbiotic star MWC 560 (V694 Mon) is an object of high interest since 1990 when it exploded and the spectra showed mass ejection with velocities around 6000 km/s [Tomov et al. 1990]. The untypical spectral behavior was accompanied by optical flickering in all colors. This flickering of MWC 560 was always present between 1984 and 2018 with a short exception during 2016. It completely disappeared in October 2018 [Goranskij et al. 2018]. V694 Mon never returned in state of rest but entered a state of continuous brightness increase modulated by different periodicities of 331 and 1860 days and possible 9570 days [Munari et al. 2016].

This study is based on observations of MWC560 during the last two observational seasons (2020/2021 and 2021/2022). Other than looking for flickering we were interested in following the variability of brightness in the same period. Looking for similarities in the spectra with other types of stars is also of great interest to us because it could help clarify the stellar configuration of such objects.

**Observations**

In this study we present CCD photometry and spectral observations of MWC560. The photometry was done by the 40-cm telescope at the Astronomical observatory of the Shumen University (AOShU) [Kjurkchieva et al. 2020]. The spectra were obtained with the ESPERO spectrograph [Bonev et al. 2017], mounted on the 2-m telescope at the Rozhen NAO.

The star was observed photometrically in two consecutive observational seasons during 2020-2022, a total of 7 nights were recorded. In Table 1 is presented the journal of these observations, all of which are done with two Sloan filters g' and r'. Exposition was 10 s with g', and 5 s and 10 s with r'.

The FWHM of the target images were around 3 pix in g′ and 4 pix in r′. The standard procedure was used for the reduction of the photometric data (de-biasing, dark frame subtraction and flat-fielding) using the software MaxImDL 6.2. We performed differential photometry using 4 standard stars. Their magnitudes (Table 2) were taken from the catalogue APASS DR9 [Henden et al. 2015]. We transformed the data using equations 2 and 3 from article [Kjurkchieva et al. 2020].

In this work we presented and analyzed two spectra of MWC560 from 6.12.2019, when it was already in a high state and flickering was absent and of XX Oph from 2019 Jun 13. Both spectra are observed with the EspeRo Echelle spectrograph [Bonev et al. 2017] on the 2.0 m telescope of the Rozhen National Astronomical Observatory, Bulgaria.

**Table 1.** Journal of observations of MWC 560 with the 40-cm telescope of the Shumen Astronomical Observatory

| Date | Band | UT start – UT end hh:mm – hh:mm | N$_{pts}$ | Average [mag] | σ [mag] | Amplitude [mag] |
|---|---|---|---|---|---|---|
| 28 November 2020 | g' | 0:12 – 3:32 | 263 | 9.096 | 0.006 | < 0.04 |
| | r' | | 264 | 8.490 | 0.007 | |
| 22 February 2021 | g' | 19:16 – 21:45 | 363 | 9.149 | 0.014 | < 0.07 |
| | r' | | 361 | 8.54 | 0.015 | < 0.08 |
| 25 April 2021 | g' | 18:30 – 20:01 | 116 | 9.072 | 0.015 | < 0.05 |
| | r' | | 116 | 8.524 | 0.011 | < 0.06 |
| 20 October 2021 | g' | 0:30 – 2:28 | 216 | 8.771 | 0.011 | < 0.05 |
| | r' | | 214 | 8.26 | 0.013 | < 0.06 |
| 08 February 2022 | g' | 18:56 – 22:30 | 307 | 8.798 | 0.017 | < 0.09 |
| | r' | | 229 | 8.351 | 0.016 | < 0.09 |
| 10 February 2022 | g' | 20:10 – 22:44 | 191 | 8.786 | 0.009 | < 0.042 |
| | r' | 20:13 – 22:23 | 154 | 8.337 | 0.008 | < 0.042 |
| 11 February 2022 | g' | 20:30 – 23:10 | 257 | 8.801 | 0.014 | < 0.05 |
| | r' | | 256 | 8.302 | 0.035 | < 0.11 |

**Table 2**. Photometrical standards used during the processing of the observations-MWC560

| Label | Star ID | RA(2000) | Dec(2000) | g' | r' |
|---|---|---|---|---|---|
| C1 | TYC 5396-491-1 | 07 26 00.69 | -07 45 34.98 | 10.644 | 10.820 |
| C2 | TYC 5396-684-1 | 07 25 27.96 | -07 44 41.84 | 9.013 | 7.913 |
| C3 | TYC 5396-916-1 | 07 25 43.28 | -07 42 04.60 | 11.771 | 10.985 |
| C4 | TYC 5396-1090-1 | 07 25 58.36 | -07 44 12.00 | 11.602 | 10.146 |
| Chk | TYC 5396-1467-1 | 07 26 16.05 | -07 50 10.14 | 10.315 | 9.360 |

**Discussion**

On 2020 November 28, for a 3.5 hours long observation run, we obtained 263 points in Sloan-g' and r' bands. During our run, no variability with amplitude larger then 0.04 mag is visible. The mean magnitudes are g'= 9.096 ± 0.006, r'= 8.490 ± 0.007. On Fig.1 is plotted the light curve together with observations from 2015 when flickering with amplitude 0.4 mag is visible. This CCD photometry of MWC 560 was performed with the 50/70 cm Schmidt telescope at the Rozhen National Astronomical Observatory.

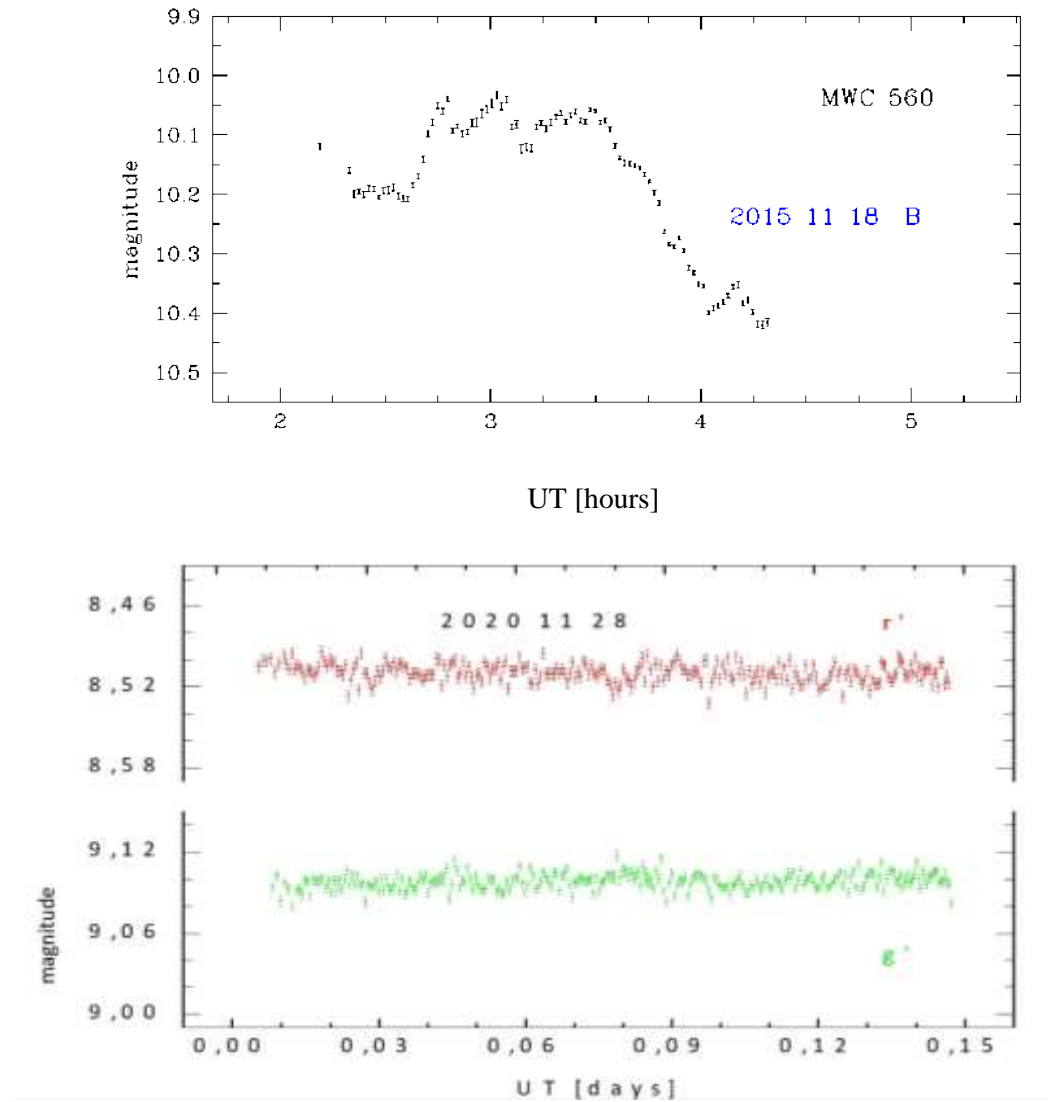

**Figure 1.** Observation MWC560 when flickering (2015, Rozhen NAO) and missing flickering (2020, AOShU)

Our first observations (2020 November 28) indicate that the optical flickering is still missing [Zamanov et al. 2020]. Our observations in 2021 are from three nights presented in Fig. 2. These observations also confirm the absence of flickering [Zamanov et al. 2021].

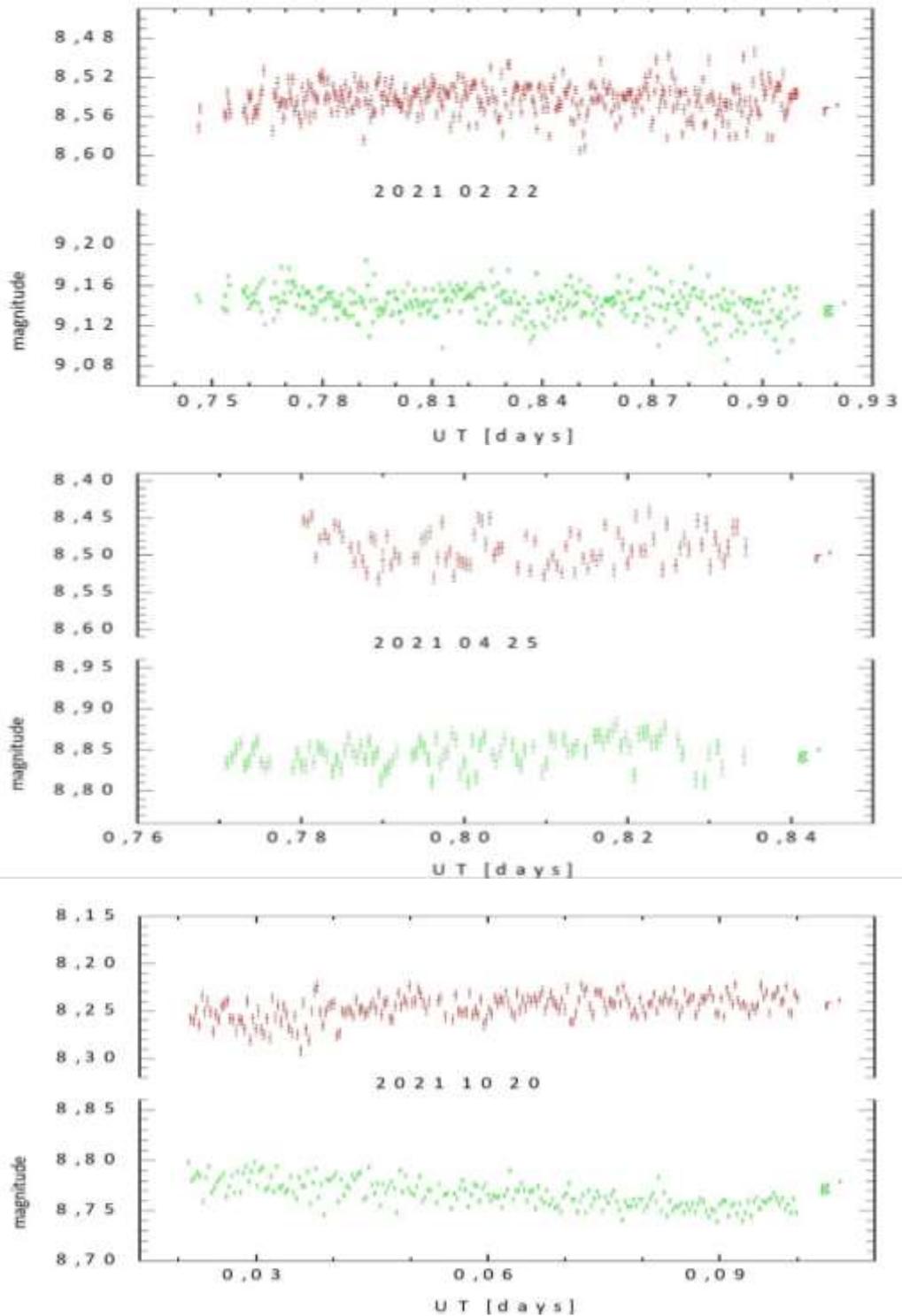

**Figure 2.** Observation MWC560 still missing flickering 2021

In 2022 so far, we have observed the star 3 times in February and the results are shown in Fig. 3. There we do not see flickering in any of the 3 nights. We were also interested in monitoring the brightness during these consecutive nights. It turned out that the brightness remains relatively high and constant, considering the estimated uncertainty ranges, but a little smaller than the maximum at October 2021. This shows a continuous high state during 2022 in agreement with the findings of [Kondratyeva et al. 2021] for the period of 2018-2021.

The increasing brightness from 2016 until 2020 shown in all spectral observations for that period suggests an expanding common shell of the system [Ando et al. 2021]. The formation of such a shell and partial or complete destruction of the accretion disk was also suggested by [Kondratyeva et al. 2021].

A study of the brightness variability of MWC560, but in a short period of time- two months, was presented by [Munari 2021]. He reports a peak of the brightness at 2021 October 27. The spectra gathered by [Goranskij et al. 2021], in November 2021 also show fast changes which is in agreement with the idea for the common shell.

Our observations show a maximum of the brightness at 2021 October 20 respectively g'=$8.771\pm0.011$ и r'=$8.26\pm0.013$, which is in agreement with the presented by [Munari 2021]. Analyzing the photometry from February 2022 we can conclude that the system is still in a high brightness state, which most likely is a result of the powerful common shell.

On Fig. 4 are plotted the spectrum of MWC 560 obtained on 2019 December 6 and the spectrum of XX Oph obtained on 2019 June 13. The spectra are almost identical. At the time when this spectrum of MWC 560 is obtained, the flickering was missing and it was probably in stage of formation of a common envelope due to the transit of the system to a dynamical mode of accretion with an increased rate [Goranskij et al. 2018].

From the observations of MWC 560 during the last three decades, it is known that it consists of a mid-M non Mira giant and an white dwarf [Lucy et al. 2020]. XX Oph is one of the two stars listed as "Iron Stars" [Bopp & Howell, 1989] due to the appearance of metal emission lines in its optical spectrum. There are no clear evidences about the components of XX Oph [Howell et al. 2009]. The spectral similarity shown in Fig. 4 is a clue that likely the components inside XX Oph are a red giant and a white dwarf. The white dwarf is probably accreting at high mass accretion rate and the components are surrounded by envelope.

This similarity can be interpreted vice versa- in this stage of the evolution of MWC560, it is mimicking behavior of an "iron star".

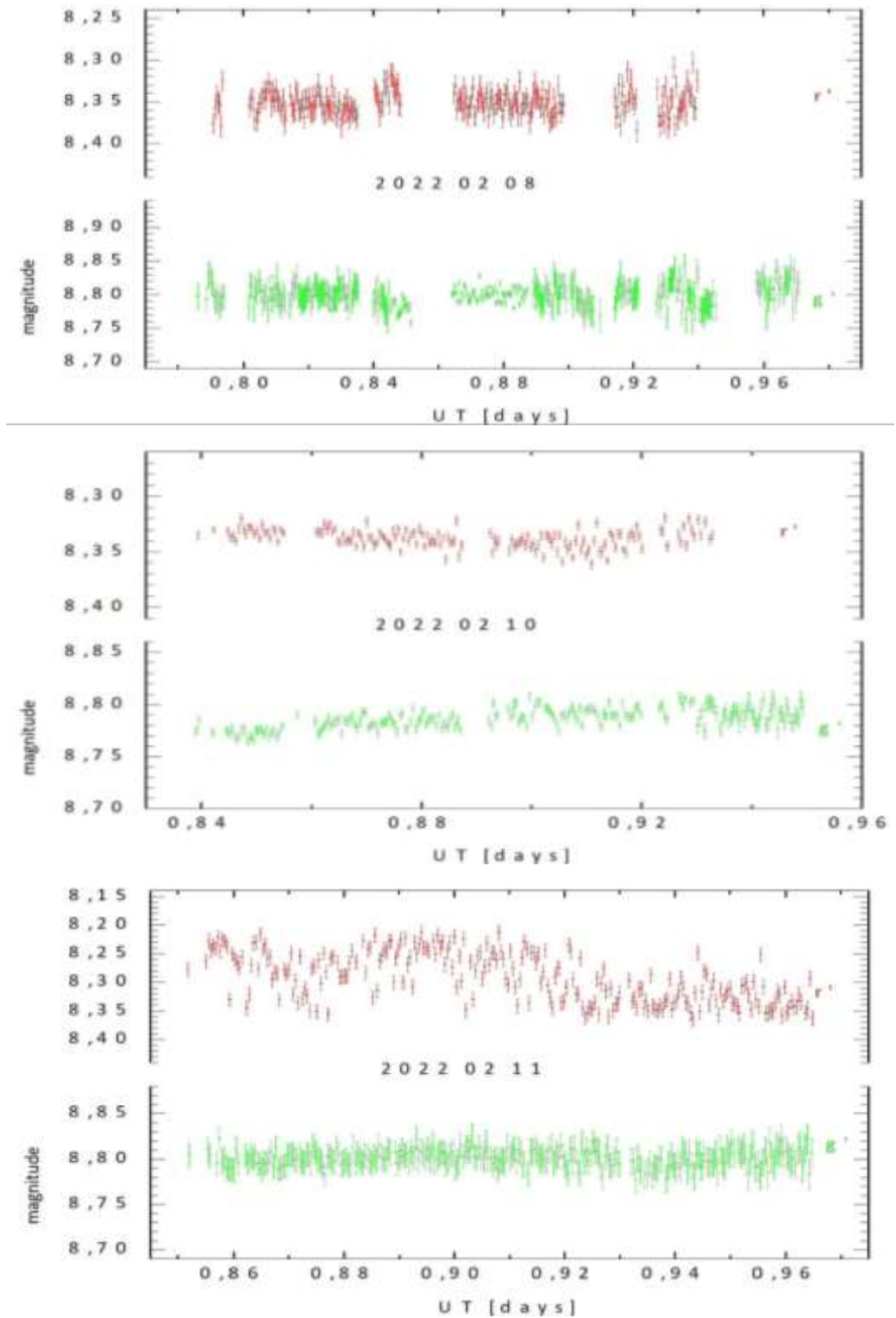

**Figure 3.** Observation MWC560 still missing flickering February 2022

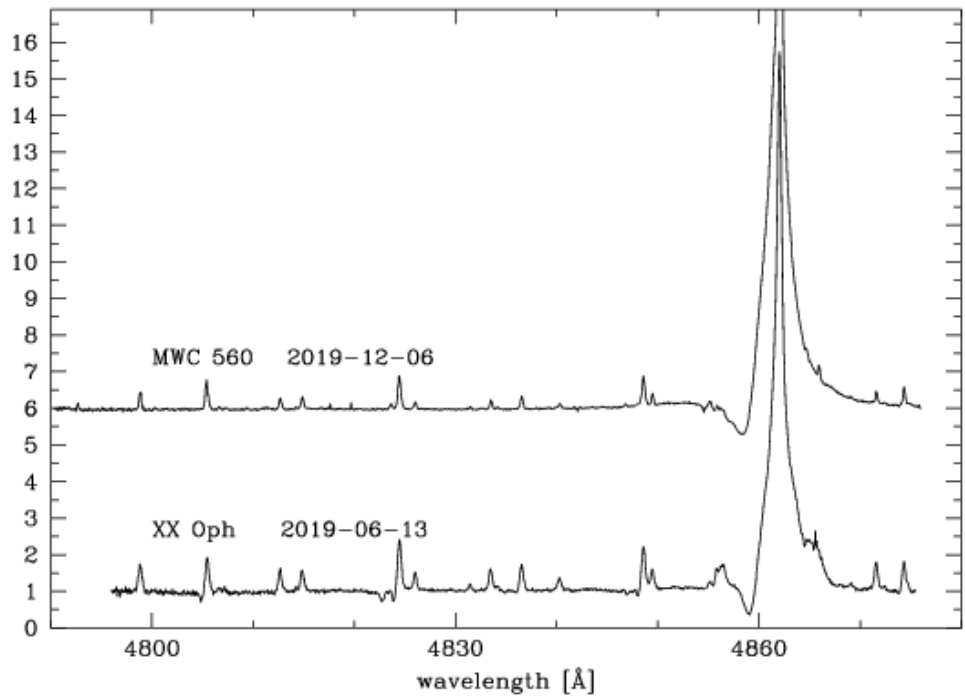

**Figure 4.** Similarity between the spectrum of MWC 560 (observed on 2019 December 6) and the spectrum of the Merrill's iron star XX Oph

**Conclusion**

Our observations during the last two observational seasons of MWC560 confirm the absence of flickering. The brightness of the system remains high, with a maximum in October 2021, also stated by [Munari 2021]. The observations from February 2022 show a high level of brightness and absence of flickering. The absence of flickering and the increased brightness are most likely caused by a wide common shell of the system which is hiding the accretion disk of the compact object.

From the similarities of the gathered spectra of XX Oph and MWC560 we assume that the components in XX Oph are a red giant and a white dwarf, which are also surrounded by a common shell.

**Acknowledgement**

This article was partially funded by projects: RD-08-100/2022 and RD-08-146/2022 of the Scientific Research Fund of Shumen University and DO-01-383/2020 (RACIO) of the Scientific Research Fund of the Bulgarian Ministry of Education and Science.